\documentclass[preprint,12pt,3p,review]{elsarticle}
\usepackage[utf8]{inputenc} 
\usepackage[T1]{fontenc}    
\usepackage{hyperref}       
\hypersetup{colorlinks,allcolors=black}
\usepackage{url}            
\usepackage{booktabs}       
\usepackage{amsfonts}       
\usepackage{nicefrac}       
\usepackage{microtype}      
\usepackage{lipsum}
\usepackage{multirow}
\usepackage{amsmath,amssymb,graphicx}
\usepackage{float}
\usepackage{relsize}

\usepackage{siunitx}
\usepackage{array}
\usepackage{lmodern}
\usepackage{amssymb}
\biboptions{sort&compress}
\usepackage{color}
\usepackage[normalem]{ulem} 
\journal{Elsevier}
\begin{document}
\begin{frontmatter}
\title{First-Principles and Reactive Molecular Dynamics Study of the Elastic Properties of Pentahexoctite-based Nanotubes}
\author[UFPI]{W. H. S. Brandão}
\author[IFPI,UNICAMP1]{J. M. De Sousa}
\ead{josemoreiradesousa@ifpi.edu.br}
\author[UFPI]{A. L. Aguiar}
\author[UNICAMP1,UNICAMP2]{D. S. Galv\~ao}
\author[UnB]{L. A. Ribeiro Junior}
\ead{ribeirojr@unb.br}
\author[UNICAMP1]{Alexandre F. Fonseca}
\ead{afonseca@ifi.unicamp.br}


\address[UFPI]{Department of Physics, Federal University of Piauí, Ininga, Teresina, 64049-550, Piauí, Brazil}


\address[IFPI]{Federal Institute of Education, Science and Technology
  of Piau\'i -- IFPI, Primavera, São Raimundo Nonato, 64770-000,
  Piauí, Brazil}
            

\address[UNICAMP1]{Applied Physics Department, Institute of Physics
  Gleb Wataghin, University of Campinas, Rua Sérgio Buarque de
  Holanda, 777 - Cidade Universitária, 13083-859, São Paulo, Brazil}


\address[UNICAMP2]{Center for Computing in Engineering and Sciences,
  University of Campinas, Rua Sérgio Buarque de Holanda, 777 - Cidade
  Universitária, Campinas, 13083-859, São Paulo, Brazil}


\address[UnB]{Institute of Physics, University of Brasília, Brasília,
  70910-900, Brazil}
            
\begin{abstract}
Pentahexoctite (PH) is a pure sp$^{2}$ hybridized planar carbon allotrope whose structure consists of a symmetric combination of pentagons, hexagons, and octagons. The proposed PH structure was shown to be an intrinsically metallic material exhibiting good mechanical and thermal stability. PH nanotubes (PHNTs) have also been proposed, and their properties were obtained from first principles calculations.
Here, we carried out fully-atomistic simulations, combining reactive (ReaxFF) molecular dynamics (MD) and density functional theory (DFT) methods, to study the PHNTs elastic properties and fracture patterns. We have investigated the mechanical properties behavior as a function of the tube diameter and temperature regimes. 
Our results showed that the PHNTs, when subjected to large tensile strains, undergo abrupt structural transitions exhibiting brittle fracture patterns without a plastic regime. 
\end{abstract}


\end{frontmatter}

\section{Introduction}

The computational design \cite{zhang2015penta,wang2015phagraphene,wang2018popgraphene,sharma2014pentahexoctite,terrones2000new,yang2013theoretical,zhang2019art,enyashin2011graphene,xu2014two,paz2019naphthylenes,lu2013two,zhuo2020me} and experimental realization~\cite{toh2020synthesis,fan2021biphenylene,hou2022synthesis,hu2022synthesis_graphyne,valentin2022} of new 2D carbon allotropes have experienced enormous growth since the advent of graphene \cite{novoselov2004electric}. The main goal guiding these studies is to obtain novel materials that overcome some electronic limitations like the known graphene null bandgap~\cite{withers2010electron} that prevent it from being fully used in nanoelectronics~\cite{tiwari2018evolution}. Recently, a 2D carbon allotrope called Pentahexoctite (PH) was proposed (see Figure \ref{fig:fig1}). It consists of a sequence of 5 -- 6 -- 8 rings of carbon atoms with sp$^{2}$ hybridization~\cite{sharma2014pentahexoctite}. Like graphene, PH is a one-atom-thick monolayer with good structural and thermal properties. It has good mechanical strength, with Young's modulus of 0.7 TPa, comparable to graphene (about 1.0 TPa \cite{jiang2009young,lee2012estimation}). 
Because of its metallic behavior and anisotropic transport properties, PH has been considered another promising structure (in terms of electronic applications) added to the increasing family of novel 2D carbon allotropes~\cite{Jana2022jap}.

It is well known that carbon atoms can be arranged throughout three hybridization patterns, i.e., sp, sp$^2$, and sp$^3$, yielding structures with very distinct topologies \cite{georgakilas2015broad,Jana2022jap}. Among these structures is worth mentioning diamond \cite{yang2019conductive}, graphite \cite{chung2002review}, amorphous structures \cite{robertson1986amorphous}, buckyballs (as C$_{60}$) \cite{kroto1991c60}, graphene \cite{geim2010rise}, and carbon nanotubes (CNTs) \cite{iijima1991helical}. In particular, the discovery of systematic synthesis of CNTs is considered an important achievement in Nanoscience, as it opened up opportunities to synthesize other nanostructures and envisioned several new applications at the nanoscale \cite{dresselhaus2000carbon,mamalis2004nanotechnology}. In this context, it is important to investigate other possible PH nanostructures, such as nanotubes. 

In the present study, the mechanical properties of PH nantubes (PHNTs) will be investigated by fully-atomistic classical molecular dynamics (MD) simulations using the state-of-the-art reactive force field (ReaxFF) and density functional theory (DFT) methods. The obtained Young's modulus values ranging between 190-354 GPa.nm are in good agreement with previous results~\cite{sharma2014pentahexoctite}. PHNTs exhibit topology-dependent mechanical properties. When subjected to a critical strain, PHNTs undergo an abrupt (brittle) transition to a completely fractured state without exhibiting a plastic deformation regime. CNTs show similar but slightly different fracture patterns. The effects of temperature on the PHNTs mechanical properties were also studied. As expected, increasing the temperature decreases the critical strain values. The PHNTs elastic properties obtained from DFT and MD simulations are in good agreement.

\section{Computational Methodology}

\subsection{Computational Modeling of PHNTs}

\begin{figure}[htb!]
    \centering
    \includegraphics[scale=0.45]{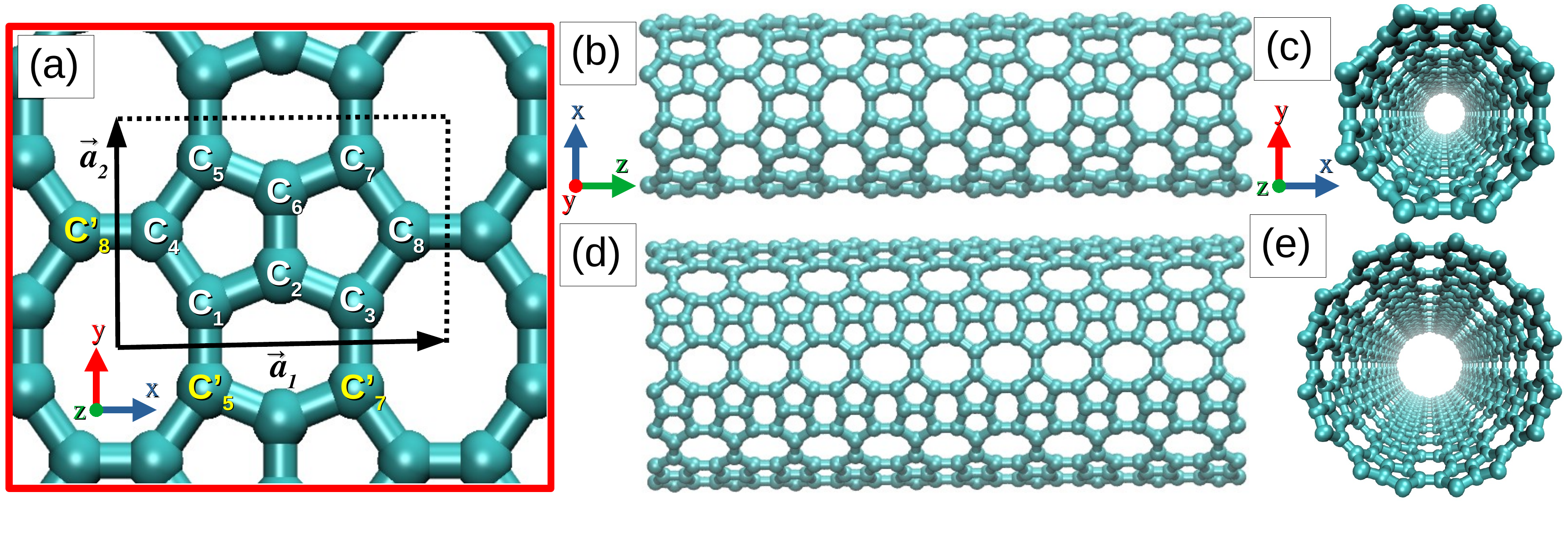}
    \caption{Schematic representation of the lattice structure of (a) PH and (b-e) PHNTs. The PH unit cell is indicated by lattice vectors $\boldsymbol{a}_1$ and $\boldsymbol{a}_2$. (b) and (c) show the lateral and frontal views of an (0,8)-PHNT, respectively. (d) and (e) show the lateral and frontal views of a (8,0)-PHNT, respectively. See the text for the meaning of $(n,m)$-PHNT nomenclature. The numbers indicated in the Figure are from DFT simulations. See text for discussions.}
    \label{fig:fig1}
\end{figure}

PH has a structure composed of rings containing 5, 6, and 8 carbon atoms, as shown in Figure \ref{fig:fig1}(a). The PH unit cell contains 8 carbon atoms, and its lattice vectors $\boldsymbol{a_1}$ and $\boldsymbol{a_2}$ are $5.85$ \r{A} and $3.78$ \r{A} long, respectively. The bond distances are C$_1$-C$_2$=1.43 \r{A}, C$_1$-C$_4$=1.49 \r{A}, C$_2$-C$_6$=1.37 \r{A}, C$_1$-C$'_5$=1.40 \r{A}, and C$_4$-C$'_8$=1.37 \r{A} (where C$'_{\mbox{i}}$ is the atom equivalent to C$_{\mbox{i}}$, see Fig. \ref{fig:fig1}).

In modeling PHNTs, we considered the same approach used to generate regular carbon nanotubes CNTs \cite{dresselhaus1998physical}. The vectors ($\boldsymbol{a}_1,\boldsymbol{a}_2$) of the PH planar unit cell can be associated with the two nanotube perpendicular vectors: $\textbf{C}_h$ and $\textbf{T}$. They are called chiral and translational vectors, respectively. The chiral vector is defined by:
\begin{equation}
\textbf{C}_h=n\boldsymbol{a}_1+m\boldsymbol{a}_2,
\label{eq:vet-ch}
\end{equation}
\noindent where $n$ and $m$ are integers that determine the lattice topology. The reduced form of the chiral vector, $\textbf{C}_h=(n,m)$, will be used. Likewise, the translational vector is given by:
\begin{equation}
\textbf{T}=t_1\boldsymbol{a}_1+t_2\boldsymbol{a}_2,
\label{eq:vet-transl}
\end{equation}
\noindent with $t_1$ and $t_2$, being also integer numbers. To identify $t_1$ and $t_2$, we apply the orthogonality condition, $\textbf{C}_h\cdot\textbf{T}=0$, with $\boldsymbol{a}_1\perp\boldsymbol{a}_2$ and $|\boldsymbol{a}_1|\neq|\boldsymbol{a}_2|$ (see Fig. \ref{fig:fig1}(a)), to obtain
\begin{equation}
    0=nt_1a^2_1+mt_2a^2_2.
    \label{eq:net-ret-t1t2}
\end{equation}
For condition \eqref{eq:net-ret-t1t2} to be satisfied, we must have
\begin{equation}
    (t_1,t_2)=\begin{cases}
    (0,1), & \mathrm{if} \quad (n,0), \\
    (1,0), & \mathrm{if} \quad (0,m).
    \end{cases}
    \label{eq:net-ret-t1t2-2}
\end{equation}

There is no chiral type nanotube ($(n,m)$ with $n\neq0$ and $m\neq0$, simultaneously) for rectangular lattice, only the zigzag type (see Fig. \ref{fig:fig1}(b)-(e)). In this case, if the chiral vector is projected onto vector $\boldsymbol{a}_1$, the translational vector will be projected onto vector $\boldsymbol{a}_2$, and vice versa.

The chiral angle $\theta$ between $\textbf{C}_h$ and $\boldsymbol{a}_1$ will be $\theta=0^\circ$ or $\theta=90^\circ$. The nanotube diameter is given by:
\begin{equation}
    d_t=\dfrac{|\textbf{C}_h|}{\pi}\Rightarrow d_t=
    \begin{cases}
    \dfrac{na_1}{\pi}, & \mathrm{if} \quad (n,0), \\
    \dfrac{ma_2}{\pi}, & \mathrm{if} \quad (0,m).
    \end{cases}
\label{eq:diam-tube}
\end{equation}

Any atom from a PH structure with a position
\begin{equation}
    \boldsymbol{R}=p\boldsymbol{a}_1+q\boldsymbol{a}_2,
    \label{eq:pos-atom}
\end{equation}
where $(p,q)$ is a pair of integers, will belong to a PHNT if the following conditions are satisfied:
\begin{eqnarray}
    0\leq\boldsymbol{R}\cdot\boldsymbol{C}_h\leq|\boldsymbol{C}_h|^2, \label{eq:cond-1}\\
    0\leq\boldsymbol{R}\cdot\boldsymbol{T}\leq|\boldsymbol{T}|^2. \label{eq:cond-2}
\end{eqnarray}

In Table \ref{tab:phognts-tab}, we present the structural parameters of the PHNTs studied here.
\begin{table}[htb!]
    \centering
    \begin{tabular}{|c|c|c|c|c|}
    \hline
         Chirality & $n$ & Number of Atoms & Diameter (\r{A}) & Length (\r{A}) \\ \hline
        \multirow{9}{*}{ $(n,0)$}&   4*   &  320, 64* & 7.45   &  37.80, 7.56* \\ \cline{2-5}
        &   5   &  400 & 9.31   &  37.80 \\ \cline{2-5}
        &   6   &  480 & 11.17   &  37.80 \\ \cline{2-5}
        &   7   &  560 & 13.03   &  37.80 \\ \cline{2-5}
        &   8*   &  640, 128* & 14.90   &  37.80, 7.56* \\ \cline{2-5}
        &   9   &  720 & 16.76   &  37.80 \\ \cline{2-5}
        &   10  &  800 & 18.62   &  37.80 \\ \cline{2-5}
        &   11  &  880 & 20.48   &  37.80 \\ \cline{2-5}
        &   12*  &  960, 192* & 22.35   &  37.80, 7.56* \\ \hline
        \hline
        \multirow{10}{*}{ $(0,n)$} &   4   &  192  & 4.81   &  35.10 \\ \cline{2-5}
        &   5   &  240 & 6.02   &  35.10 \\ \cline{2-5}
        &   6*   &  288, 96* & 7.22   &  35.10, 11.70* \\ \cline{2-5}
        &   7   &  336 & 8.42   &  35.10 \\ \cline{2-5}
        &   8   &  384 & 9.63   &  35.10 \\ \cline{2-5}
        &   9   &  432 & 10.83  &  35.10 \\ \cline{2-5}
        &   10  &  480 & 12.03  &  35.10 \\ \cline{2-5}
        &   11  &  528 & 13.24  &  35.10 \\ \cline{2-5}
        &   12* &  576, 192*    & 14.44   &  35.10, 11.70* \\ \hline
    \end{tabular}
    \caption{Structural parameters of the model PHNTs. The tube chiralities considered here are $(0,n)$ and $(n,0)$. Superscript asterisk (*) indicates the nanotubes that are also simulated with DFT.}
    \label{tab:phognts-tab}
\end{table}

In the next subsections, the computational details of the MD simulations and DFT calculations of the structural and mechanical properties of PHNts will be shown.

\subsection{Reactive (ReaxFF) MD Simulations}

Classical MD simulations with the reactive force field (ReaxFF) were performed using the LAMMPS code \cite{plimpton1995fast}. ReaxFF \cite{van2001reaxff} is an interatomic potential that allows the breaking and formation of chemical bonds. This reactive feature is necessary for a good description of the mechanical properties beyond the linear regime, including plastic deformation and fracture. The parameters considered in the ReaxFF potential as used here are described in Refs. \cite{mueller2010development,van2001reaxff}. The molecular visual dynamics (VMD) software \cite{vmd96} was used to build the structural MD snapshots.

The PHNTs were subjected to uniaxial stresses along their longitudinal direction in the MD simulations. Stress-strain curves were obtained for temperature values ranging from 10 up to 1200 K. 

The integration of the Newton's equations of motion were performed using the velocity Verlet algorithm \cite{martys1999velocity}. The time step of 0.05 fs was used in all MD simulations. PHNT structures were equilibrated using Nose-Hoover thermostats \cite{evans1985nose} for a total of 10000 fs before the stretching dynamics. Residual lattice stresses in the thermalized PHNTs were eliminated by performing an additional NPT simulation \cite{andersen1980molecular}, with pressure set to zero, for more 10000 fs. The mechanical properties (both elastic and fracture patterns) of PHNTs at different temperatures were investigated by stretching the structures at a constant engineering tensile strain rate of $10^{-6}/$fs. These protocols have been successfully used to reveal the mechanical properties of several  1D~\cite{NairBuehler2011carbyne,KocsisCranford2014carbyne,DESOUSA201614,DESOUSA2019109153,brandao2021PopG} and 2D~\cite{ROMAN201913,DESOUSA2021111052,SHISHIR2021103895,brandao2022,Brandao2023MechMat} nanomaterials.  

\subsection{DFT simulations}

The DFT calculations were carried out using the SIESTA code \cite{soler2002}. Generalized gradient approximation (GGA) with the exchange-correlation potential PBE \cite{pbe1996} was considered in the present study. An $8\times8\times1$ Monkhorst-Pack \cite{mpack1976} grid was considered in the self-consistent calculations. The double-$\zeta$ polarization (DZP) basis set was used, 
as well as 0.05 and 400 Ry for orbital cutoff \cite{pao2001} and the mesh cutoff, respectively, were considered \cite{anglada2002}. 

DFT was also employed in the study of the PHNT uniaxial stretching. The PHNT supercells were deformed from 1 up to 30\% strain along their longitudinal direction. The PHNTs were optimized using the tolerance for the force component of 0.01 eV/\r{A} at each strain step. As usual, periodic boundary conditions were considered along the tube longitudinal direction. The vacuum distances along the x- and y-directions are equal to 40 \r{A}, which is more them twice the diameter of most of the tubes, so enough to avoid spurious effects due to the lattice images. The PHNTs studied with DFT are: (4,0), (8,0), (12,0), (0,6), and (0,12), as highlighted in Table \ref{tab:phognts-tab}. 

\section{Results and Discussion}

\begin{figure}[htb!]
    \centering
    \includegraphics[scale=0.5]{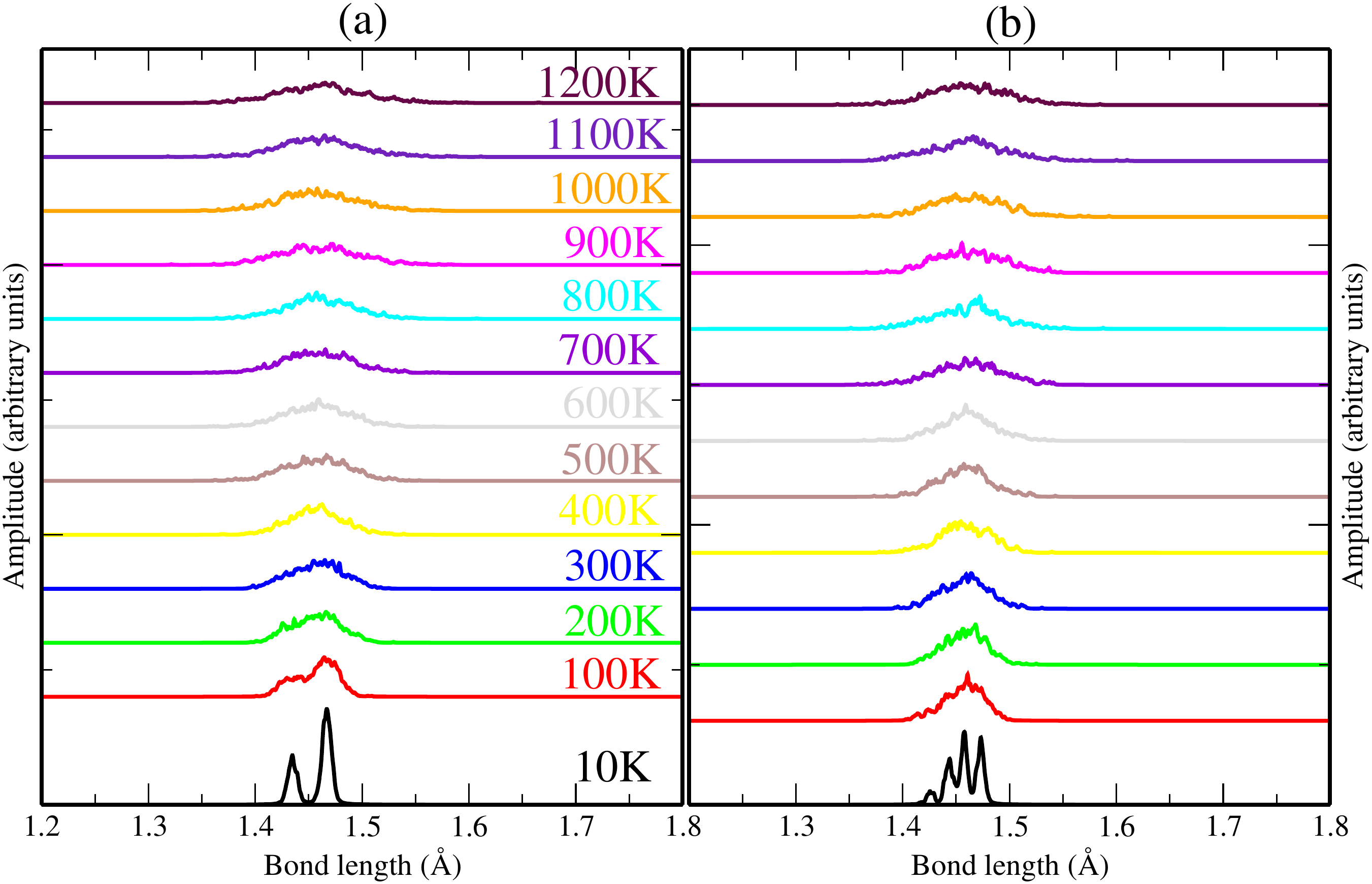}
    \includegraphics[scale=0.5]{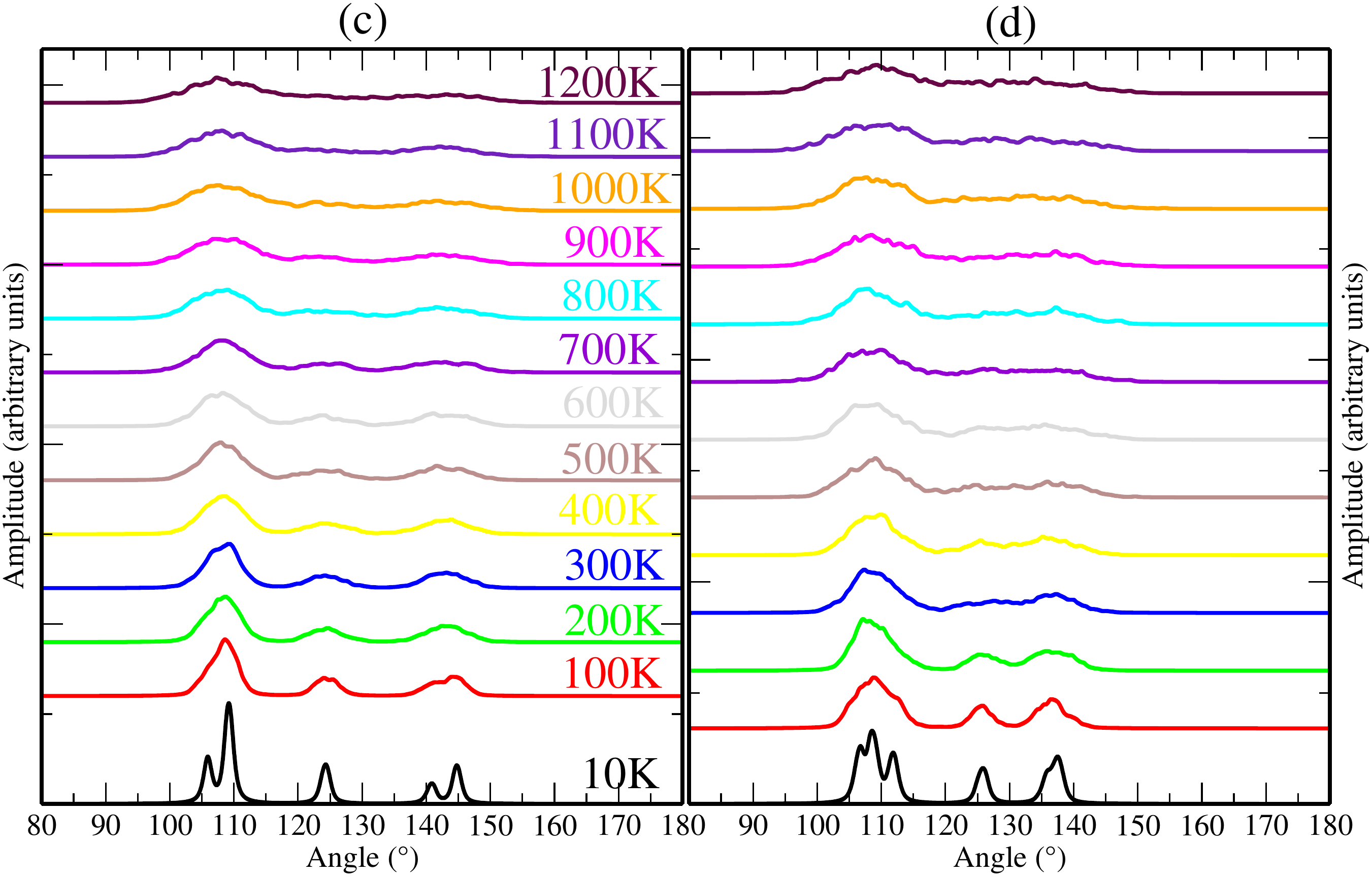}
    \caption{Bond length and bond angle distributions for (8,0) in (a) and (c), and  (0,8) in (b) and (d) PHNTs, respectively. The results are for all simulated temperatures and normalized to the values at 10 K.}
    \label{fig:bon-ang-dist}
\end{figure}

In Figure \ref{fig:bon-ang-dist}, we present the distributions of bond length and bond angle values of the (8,0) and (0,8) PHNTs for 10 K < \textit{T} < 1200 K. Different bond lengths and angles values are distinguishable only for very low temperature.

\begin{figure}[htb!]
    \centering
    \includegraphics[scale=0.5]{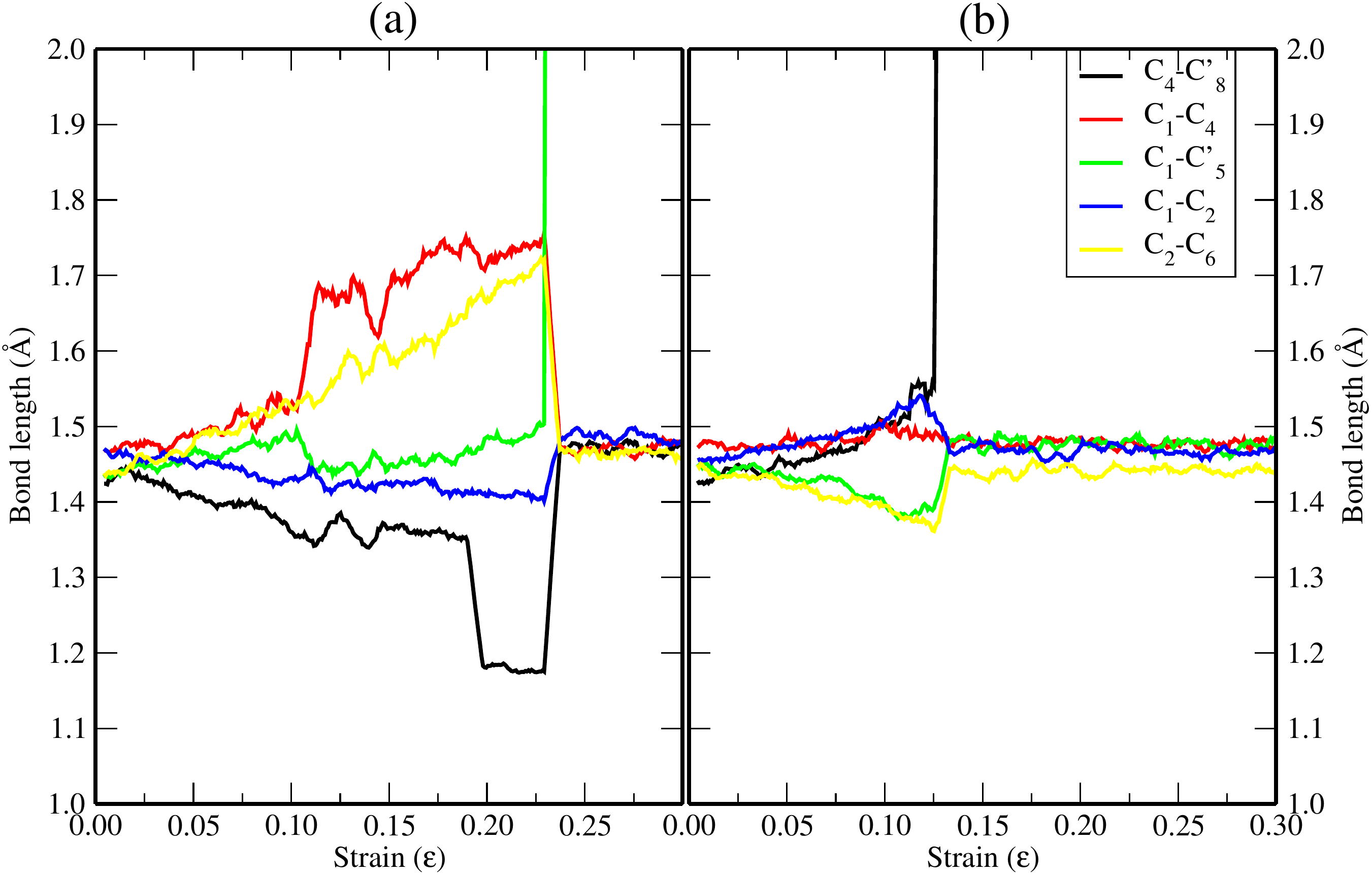}
    \includegraphics[scale=0.5]{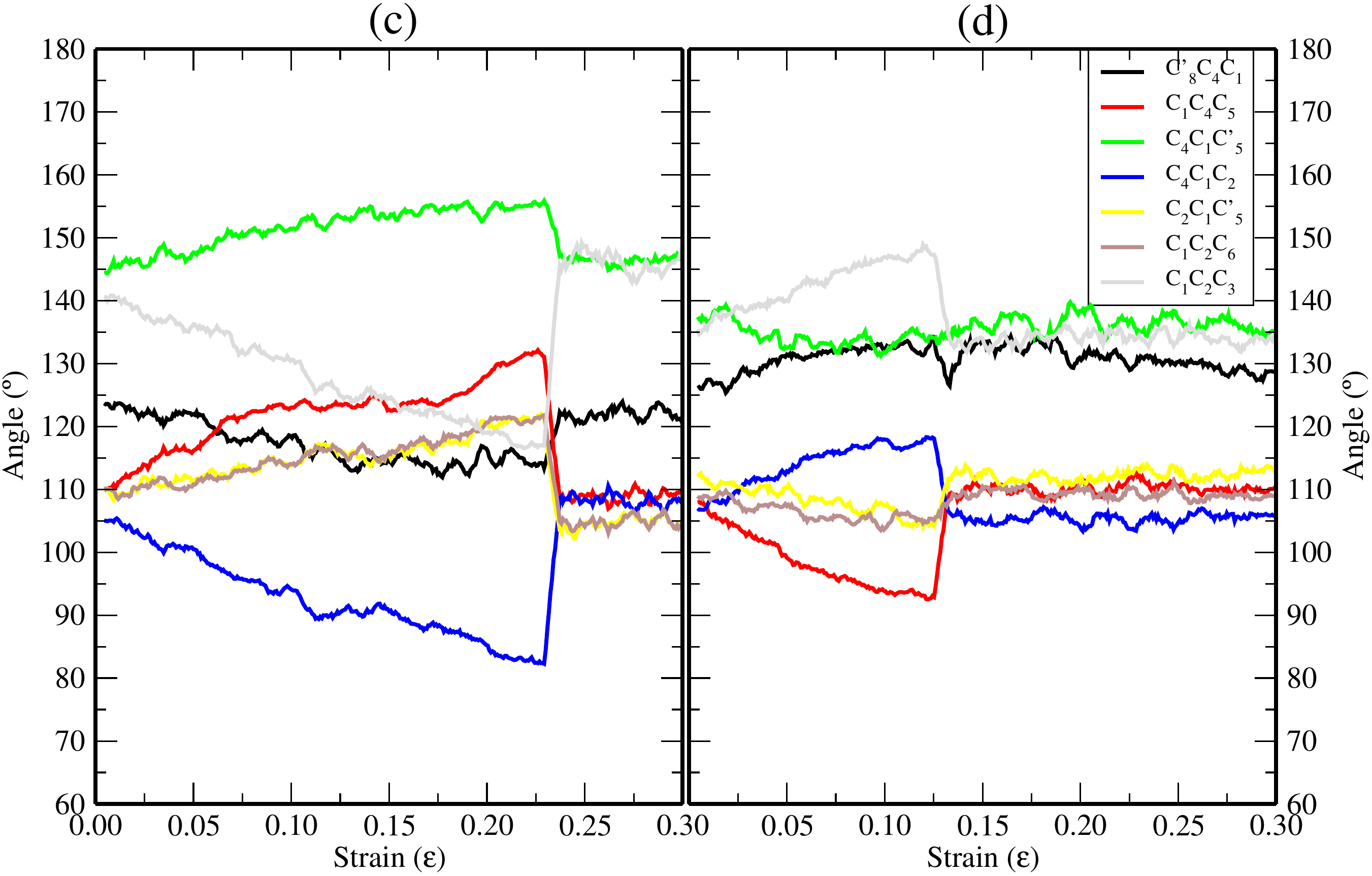}
    \caption{Bond length and bond angle evolution with strain for (8,0) in (a) and (c), and (0,8) in (b) and (d) PHNTs, respectively, at room temperature. In Figure \ref{fig:fig1}a is indicated the labeling of each C$_{\mbox{i}}$ carbon atom.}
    \label{fig:bond-ang-evol}
\end{figure}

In Figure \ref{fig:bond-ang-evol}, we present some selected bond lengths and bond angles from (8,0) and (0,8) PHNTs as a function of the applied tensile strain, from the fully relaxed structure up to its fracture. In Figure \ref{fig:fig1}a, it is indicated the labeling of each C$_{\mbox{i}}$ carbon atom shown in Figure \ref{fig:bond-ang-evol}. It is possible to see that some carbon-carbon (C-C) bond distances increase while others decrease or remain practically constant. One interesting observation for the (8,0) PHNT is that the C-C bonds that are broken are not necessarily the ones that are more strained. For the (0,8) PHNT, on the other hand, the C-C bonds that break are the ones that are more strained. Another interesting feature is that during the tensile strain of both tube types, there are C-C bonds that decrease in length. These bonds are those that are orthogonal to the tensile strain direction. The length decrease of these bonds reflects the Poisson effect, i.e., the tendency to decrease the diameter size when the tube is tensile strained.

The behavior of the bond angles is also relevant to understand the deformation of PHNTs. For both the (8,0) and (0,8) PHNTs, the bond angles for C-C bonds that are roughly aligned (orthogonal) to the tensile strain increased (decreased) with the amount of strain. 

\begin{figure}[htb!]
    \centering
    \includegraphics[width=\linewidth]{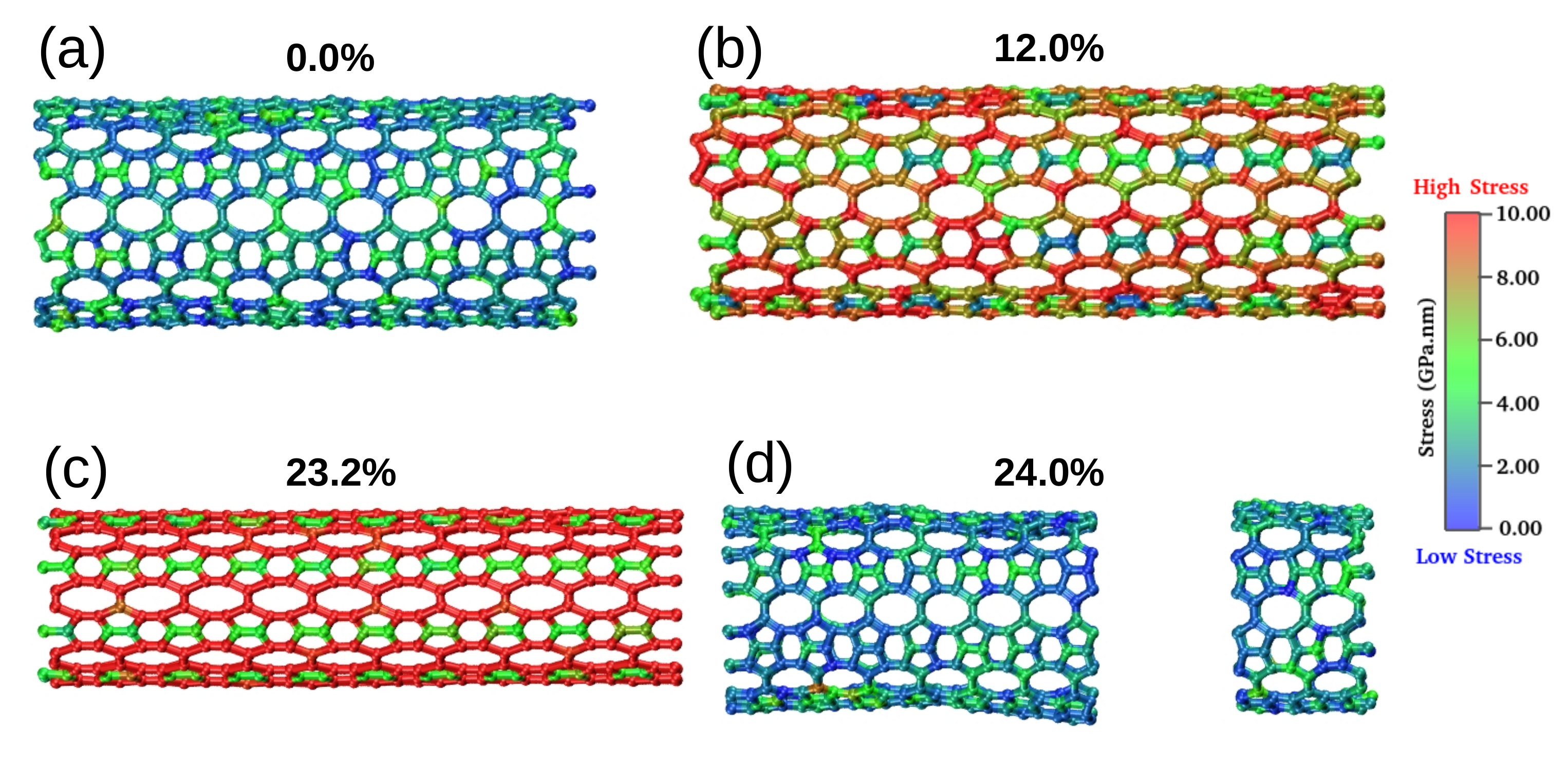}
    \caption{Representative MD snapshots for the the fracture process of (8,0)-PHNT at 300K. Panels (a), (b), (c), and (d) show the bond arrangements for the nanotube when it is subjected to a strain of 0.0 \%, 12.0 \%, 23.2 \%, and 24.0 \%, respectively.}
    \label{fig:fig2}
\end{figure}

\begin{figure}[htb!]
    \centering
    \includegraphics[width=\linewidth]{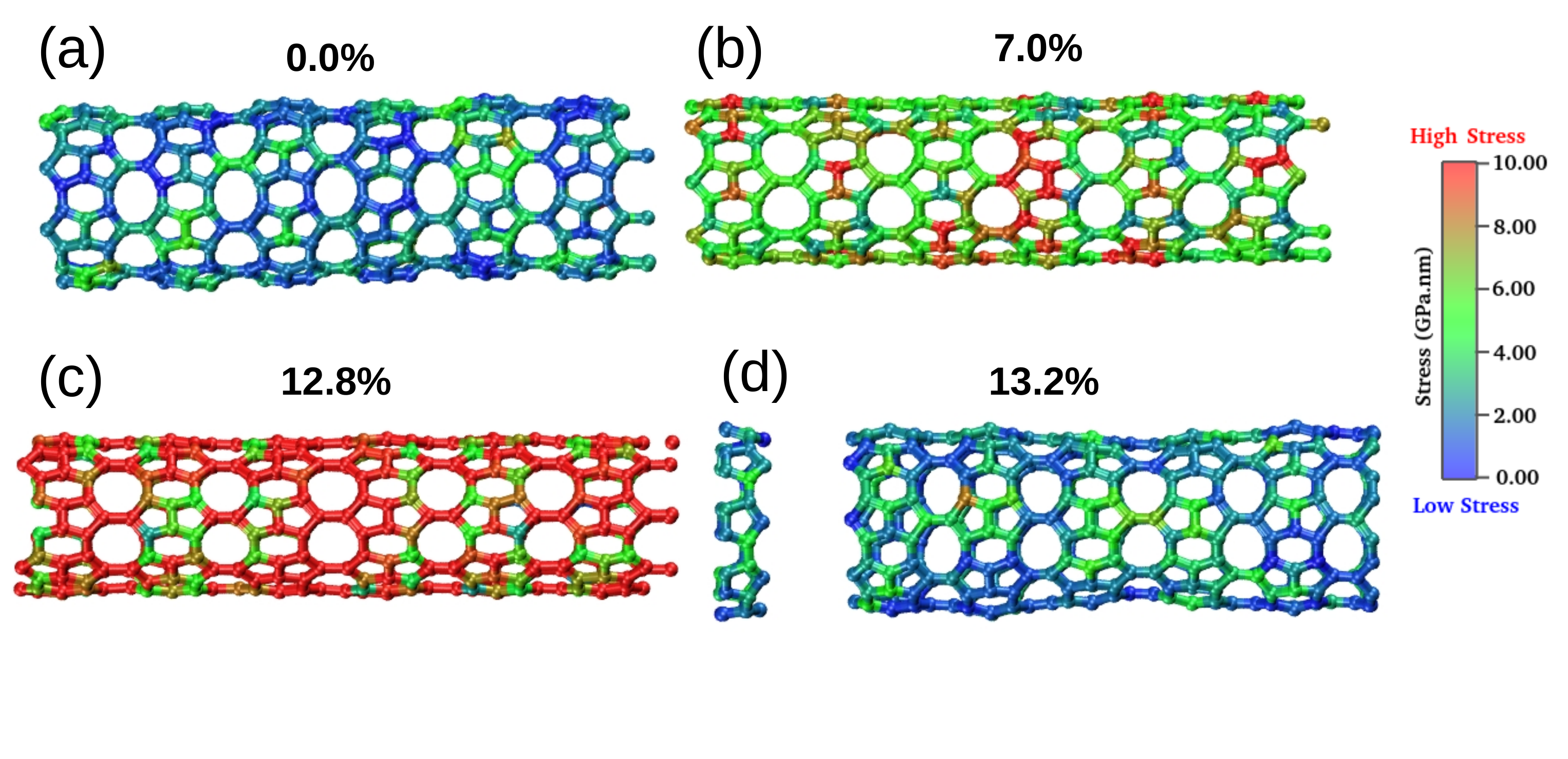}
    \caption{Representative MD snapshots for the the fracture process of (0,8)-PHNT at 300K. Panels (a), (b), (c), and (d) show the bond arrangements for the nanotube when it is subjected to a strain of 0.0 \%, 7.0 \%, 12.8 \%, and 13.2 \%, respectively.}
    \label{fig:fig3}
\end{figure}

Figures \ref{fig:fig2} and \ref{fig:fig3} show MD snapshots of the fracture process of (8,0) and (0,8) PHNTs at 300 K, respectively. The spatial distribution of von Mises (VM) stress per-atom values \cite{mises_1913} are shown based on a color scheme (from red/large stress to blue/low stress) in these figures. 
The VM stresses allow to find out  potential fracture points or regions. Computational details regarding the VM calculations can be found in Ref. \cite{felix2020mechanical}. 

In Figures \ref{fig:fig2} and \ref{fig:fig3}, we present MD snapshots of the tensile strained (8,0)- and (0,8)-PHNTs, respectively. From Figure \ref{fig:fig2}, we can see that when the (8,0)-PHNT is subjected to a critical strain (Fig.~\ref{fig:fig2}~(c)), it undergoes an abrupt structural transition to a fractured state. This process occurs without plastic-like stages, linear chain formation, or ring reconstructions. This result is quite interesting because the original two-dimensional Pentahexaoctite exhibits the appearance of linear atomic chains just before fracture when under tensile strain \cite{brandao2022}. The complete absence of a plastic regime in PHNTs is unusual since regular \cite{ragab2009} and amorphous CNTs~\cite{junior2021reactive} present plastic deformations before becoming completely fractured. 
Figure \ref{fig:fig2}(a) shows the (8,0)-PHNT at 0.0 \% of strain. At 12 \% of strain, the tensile load begins to distribute stress throughout the nanotube in a non-uniform pattern, as shown in Figure \ref{fig:fig2}(b). Figure \ref{fig:fig2}(c) shows the structure at the critical strain of 23.2 \%, just before fracture. We can see that although all the lattice rings were stretched, some of the C$-$C bonds are less stretched than the others (green color regions). At this threshold, the stress is uniformly distributed along the atoms of the eight-membered rings (highly-stressed, red color regions). The C$-$C bonds connecting the six-membered rings of carbon atoms tend to remain unstressed at the cost of changing the bond angles of the eight-membered rings. It favors Poisson's effect, i.e., the decrease of the PHNT diameter under tensile strain. 
 
In the (8,0)-PHNT fracture process, the breaking bonds are just the ones that laterally connect the pairs of pentagons and that are aligned to the direction of the applied strain. Curiously, the C-C bonds from the five-membered rings, which are also aligned with the tensile strain direction, did not experience much stress during the tensile strain. This feature is a consequence of a more compliant C$^{'}_8$-C$_4$-C$_1$ bond angle. In Figure \ref{fig:fig2}(d), one can observe that once the structure of the (8,0)-PHNT gets fractured, the separated fragments recover their original morphologies, which indicates that the accumulated strain was fully elastic, without the formation of a plastic regime. This behavior is important to applications where mechanical energy can be storaged at the nanoscale because the structure can retain large amounts of tensile strain (elastic energy) without damage until the limit where it is fractured.

The representative (0,8)-PHNT case shows a similar fracture process, as depicted in Figure \ref{fig:fig3}. (0,8)-PHNT also undergoes an abrupt structural transition to a fractured stage just after the critical strain. The associated broken bonds are those that longitudinally connect the pair of pentagons and that are aligned to the direction of the applied strain. There are neither ring reconstructions nor linear atomic chain formations. At 7 \% of strain, the tensile load distributes the stress quite uniformly (except for local thermal fluctuations) throughout the nanotube, as shown by Figure \ref{fig:fig3}(b). 

\begin{figure}[htb!]
    \centering
    \includegraphics[width=\linewidth]{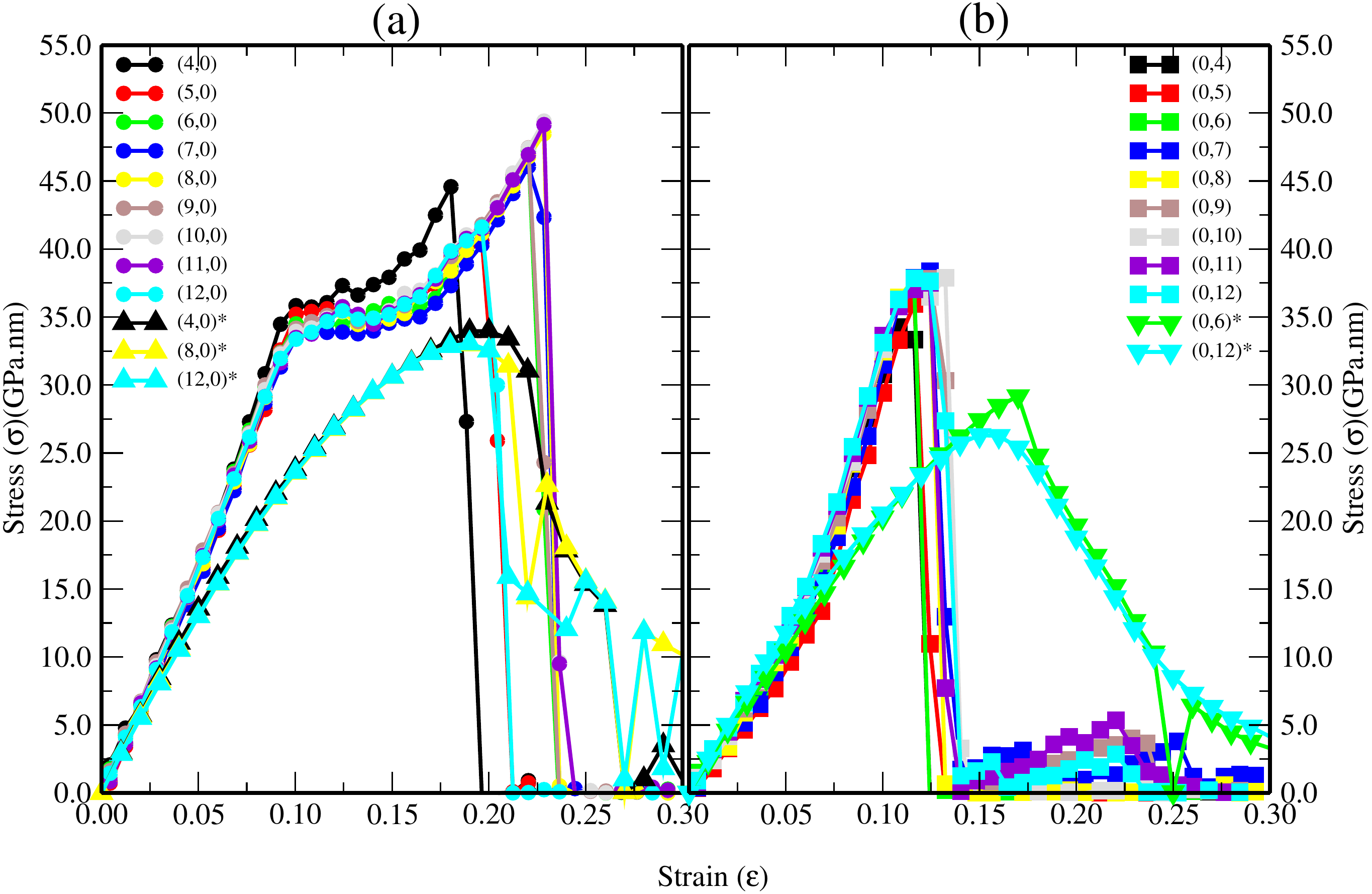}
    \caption{Stress-strain curves for 300 K for all (a) $(n,0)$-PHNTs and (b) $(n,0)$-PHNTs studied here. Circles and squares refer to the MD results while triangles refer to DFT ones.}
    \label{fig:fig4}
\end{figure}

The (0,8)-PHNT critical strain value, about 12.8 \%, is about half that of the (8,0)-PHNT. This can be understood in terms of the straining mechanism of the structure. Inspection of the variation of the pairs of bond angles defined by C$_1$-C$_4$-C$_5$ and C$_7$-C$_8$-C$_3$ (C$_1$-C$_2$-C$_3$, and C$_5$-C$_6$-C$_7$) atoms with the applied tensile strain on (8,0)-PHNT ((0,8)-PHNT) shows their significant role on the PHNTs strain mechanism. In (8,0)-PHNT ((0,8)-PHNT), 
the C$_1$-C$_4$-C$_5$ and C$_7$-C$_8$-C$_3$ (C$_1$-C$_2$-C$_3$ and C$_5$-C$_6$-C$_7$) bond angles belong to the 5-membered ring (6-membered ring) of the structure, and their corresponding bonds are roughly aligned with the tension direction. When tensile strained, these bond angles significantly change. Figures~\ref{fig:bond-ang-evol} (c) and (d) confirm that. When tensile straining the (8,0)-PHNT ((0,8)-PHNT), C$_1$-C$_4$-C$_5$ increases (decreases) and C$_1$-C$_2$-C$_3$ decreases (increases). As a consequence, the 8-membered rings in (8,0)-PHNTs and 6-membered rings in (0,8)-PHNTs become stretched along the tensile strain direction. As the 8-membered ring can accumulate larger strain values than the 6-membered one, the critical strain of ($n$,0)-PHNTs is expected to be larger than that of (0,$n$)-PHNTs. That is exactly what we see in Figure \ref{fig:fig4}. In fact, there are more C-C bonds perpendicular to the applied strain direction in (0,$n$)-PHNTs than in ($n$,0)-PHNTs. Those perpendicular bonds will
not contribute to the mechanical resistance to tensile strain. Finally, as occurs to the (8,0)-PHNT, Figure \ref{fig:fig3}(d) shows that after the (0,8)-PHNT is fractured, it also recovers its original morphology, which again shows that the accumulated strain was fully elastic, without a plastic regime.  

Figures \ref{fig:fig4}(a) and \ref{fig:fig4}(b) show the stress-strain curves, obtained from MD and DFT simulations, for (0,$n$)-PHNTs and ($n$,0)-PHNTs at 300 K, respectively. The (0,$n$)$^{*}$ and ($n$,0)$^{*}$ cases correspond to the DFT results. These results confirm the already predicted~\cite{sharma2014pentahexoctite} fact 
that PHNTs have chirality-dependent 
mechanical properties. As happened to the PHNTs with $n=8$, $(n,0)$-PHNTs are more resilient to tensile loading than $(0,n)$-PHNTs. The critical strain values of $(n,0)$-PHNTs are always larger than the corresponding ones of $(0,n)$. It happens because, during the tensile strain, the 8-membered rings in $(n,0)$-PHNTs can accumulate much more strain than the 6-membered rings in $(0,n)$-PHNTs. It should be stressed that in $(n,0)$-PHNTs, there are more C$-$C bonds aligned with the direction of the tensile strain than $(0,n)$-PHNTs. If the aligned C-C bonds are considered as a serial combination of springs, the larger the number of springs, the larger the accumulated deformation. In this sense, $(n,0)$-PHNTs are expected to present higher ultimate strength (US), 
critical strain ($\varepsilon_C$), and later fracture or critical stress ($\sigma_C$) than $(0,n)$-PHNTs. It is worth mentioning that the US is defined as the maximum tensile stress held by the structure, and in some cases, it coincides with the critical stress, $\sigma_C$, at the moment of fracture. $\varepsilon_C$ denotes the strain value at the moment of the fracture. Virial stresses, $\sigma$, along the stretching direction were calculated as  described in reference \cite{de2021nanostructures}.

\begin{figure}[htb!]
    \centering
    \includegraphics[width=\linewidth]{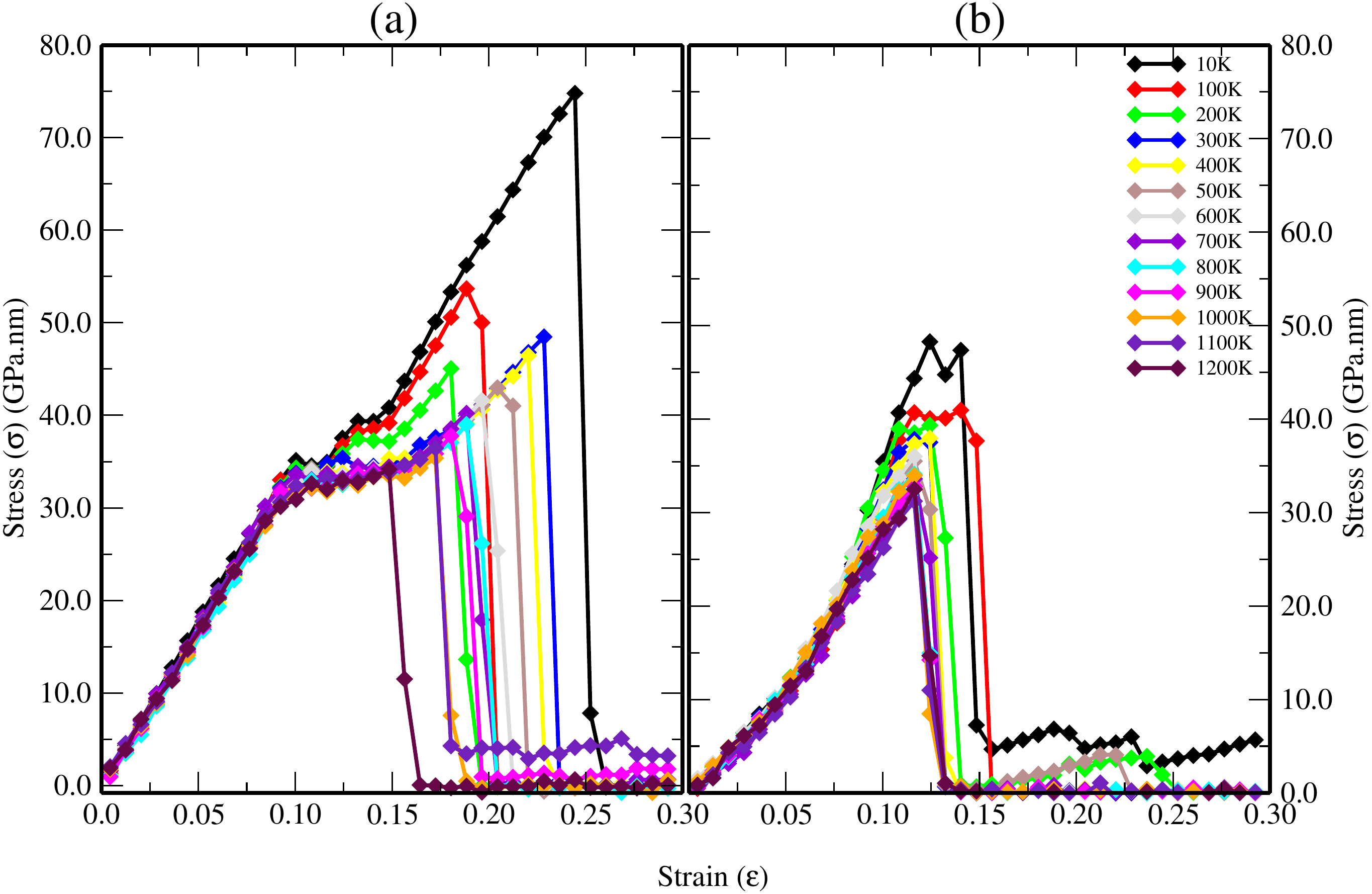}
    \caption{Stress-strain curves at 300 K for all (a) (8,0)-PHNT and (b) (0,8)-PHNT. Circles and triangles refer to the MD and DFT points, respectively.}
    \label{fig:fig5}
\end{figure}

\begin{figure}[htb!]
    \centering
    \includegraphics[scale=0.8]{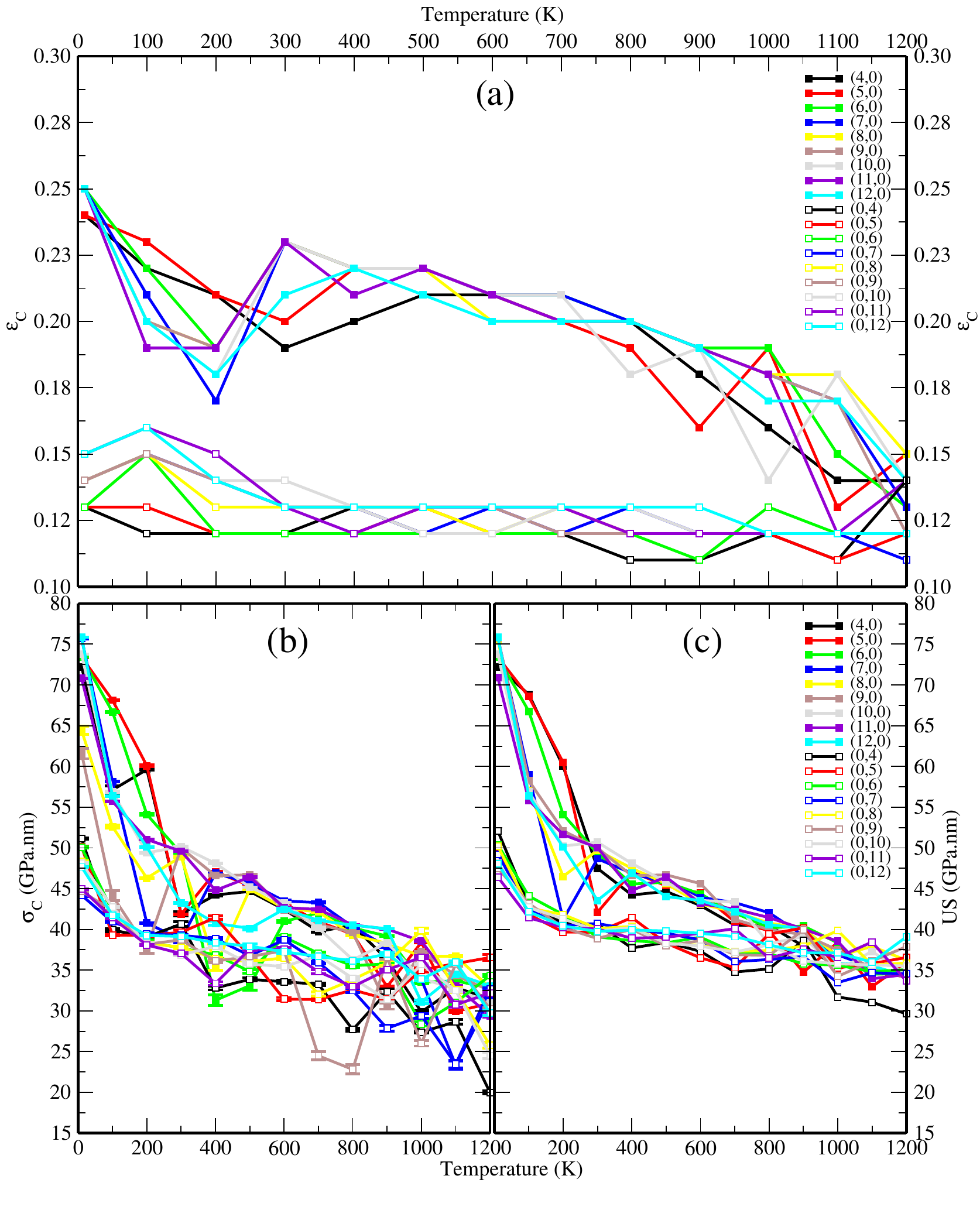}
    \caption{(a) $\varepsilon_C$, (b) $\sigma_C$ and (c) US as a function of the temperature.}
    \label{fig:ss-temp}
\end{figure}

The temperature dependence of the mechanical properties of PHNTs was also investigated. Figures \ref{fig:fig5}(a) and \ref{fig:fig5}(b) show the stress-strain curves for the (8,0)-PHNT and (0,8)-PHNT for representative cases, respectively. In general, the observations made earlier for the differences between the stress-strain curves of $(n,0)$-PHNTs and $(0,n)$-PHNTs still hold. Furthermore, we can see that, in general, the higher the temperature, the lower the critical UC, $\sigma_C$, and $\varepsilon_C$ values of all PHNTs. The profile of the curves in Figures \ref{fig:fig5}(a) and \ref{fig:fig5}(b) is similar among the same class of nanotubes regardless of the temperature value. Except for few values of temperatures, UC and $\sigma_C$ coincide. This result is consistent with the absence of a plastic regime of deformation of the PHNTs before fracture. There is, however, one remarkable difference in the temperature dependence of the critical strain, $\varepsilon_C$, from 300 to 1200 K. For (8,0)-PHNTs, the critical strain decreased by about 30 \% while for (0,8)-PHNTs it decreased by only about 4 \%. Figure \ref{fig:ss-temp}(b) confirms this trend for all $(n,0)$-PHNTs investigated here. It seems that $(n,0)$-PHNTs become stiffer at higher temperatures. Table \ref{tab:table2} presents the values obtained for the elastic properties of selected $(n,0)$-PHNT and $(0,n)$-PHNT for all temperature values simulated here.

\begin{table}[htb!]
\centering
\caption{Mechanical properties values for PHNTs. The linear regime considered was 5\% of strain.}
\label{tab:table2}
\begin{tabular}{|c|c|c|c|c|c|}
\hline
TYPE & TEMP. (K) & $Y_{mod}$ (GPa.nm) & US (GPa.nm) & $\sigma_C$ (GPa.nm) & $\varepsilon_C$ \\ \hline
\multirow{5}{*}{(4,0)}  & 10 & 345.29 $\pm$   1.74 & 72.22 $\pm$   0.13 & 72.22 $\pm$  0.13 & 0.24 \\ \cline{2-6}
& 300   &   324.01 $\pm$   6.68  & 47.49 $\pm$   0.37  & 41.93 $\pm$  0.79  &  0.19    \\ \cline{2-6}
& 600   &   303.62 $\pm$   9.49  & 42.92 $\pm$   0.13  & 42.47 $\pm$  0.13  &  0.21    \\ \cline{2-6}
& 900   &   300.82 $\pm$  11.84  & 37.95 $\pm$   0.52  & 36.30 $\pm$  0.30  &  0.18    \\ \cline{2-6}
& 1200  &   305.74 $\pm$  12.56  & 34.08 $\pm$   0.72  & 31.55 $\pm$  0.44  &  0.14    \\ \hline
\multirow{5}{*}{(12,0)} & 10 & 353.00 $\pm$   1.29 & 75.86 $\pm$   0.11 & 75.86 $\pm$  0.11 & 0.25 \\ \cline{2-6}
& 300   &   326.27 $\pm$   3.69  & 43.55 $\pm$   0.23  & 43.23 $\pm$  0.23  &  0.21    \\ \cline{2-6}
& 600   &   329.79 $\pm$   4.98  & 43.55 $\pm$   0.27  & 42.50 $\pm$  0.26  &  0.20    \\ \cline{2-6}
& 900   &   316.85 $\pm$   6.92  & 40.09 $\pm$   0.15  & 40.09 $\pm$  0.15  &  0.19    \\ \cline{2-6}
& 1200  &   311.67 $\pm$   8.24  & 34.31 $\pm$   0.31  & 33.07 $\pm$  0.29  &  0.14    \\ \hline
(4,0)  & \multirow{3}{*}{DFT} & 270.59 $\pm$ 4.02 &  33.97 $\pm$ 0.76 &  33.97 $\pm$ 0.76 & 0.20  \\ \cline{1-1} \cline{3-6}
(8,0)  & & 259.61 $\pm$ 5.29 &  33.00 $\pm$ 0.44 &  33.00 $\pm$ 0.44 & 0.19  \\ \cline{1-1} \cline{3-6}
(12,0) & & 258.66 $\pm$ 3.68 &  33.07 $\pm$ 4.30 &  32.55 $\pm$ 6.11 & 0.20  \\ \hline
\multirow{5}{*}{(0,4)}  & 10 & 211.27 $\pm$   1.38 & 52.07 $\pm$   0.31 & 51.15 $\pm$  0.27 & 0.13 \\ \cline{2-6}
& 300   &   199.71 $\pm$   8.60  & 40.64 $\pm$   0.95  & 40.64 $\pm$  0.95  &  0.12    \\ \cline{2-6}
& 600   &   190.83 $\pm$  11.69  & 37.30 $\pm$   0.82  & 33.57 $\pm$  0.66  &  0.12    \\ \cline{2-6}
& 900   &   202.83 $\pm$  14.02  & 38.69 $\pm$   1.28  & 32.32 $\pm$  1.27  &  0.11    \\ \cline{2-6}
& 1200  &   224.96 $\pm$  17.38  & 29.63 $\pm$   1.11  & 19.96 $\pm$  0.64  &  0.14    \\ \hline
\multirow{5}{*}{(0,12)} & 10 & 239.80 $\pm$   1.61 & 48.08 $\pm$   0.20 & 47.67 $\pm$  0.20 & 0.15 \\ \cline{2-6}
& 300   &   237.22 $\pm$   6.05  & 39.75 $\pm$   0.61  & 39.14 $\pm$  0.28  &  0.13    \\ \cline{2-6}
& 600   &   252.90 $\pm$   8.13  & 39.50 $\pm$   0.69  & 37.24 $\pm$  0.62  &  0.13    \\ \cline{2-6}
& 900   &   226.62 $\pm$   8.62  & 37.11 $\pm$   0.60  & 36.99 $\pm$  0.58  &  0.13    \\ \cline{2-6}
& 1200  &   238.64 $\pm$   8.40  & 39.08 $\pm$   0.98  & 29.76 $\pm$  1.00  &  0.12    \\ \hline
(0,6)  & \multirow{2}{*}{DFT} & 207.94 $\pm$ 4.33 &  29.17 $\pm$ 2.18 &  29.17 $\pm$ 2.18 & 0.17  \\ \cline{1-1} \cline{3-6}
(0,12) & & 236.41 $\pm$ 3.90 &  26.30 $\pm$ 0.50 &  26.30 $\pm$ 0.50 & 0.15  \\ \hline
\end{tabular}
\end{table}

The temperature dependence of the critical stress, critical strain, and US values are shown in Figures \ref{fig:ss-temp}(a), \ref{fig:ss-temp}(b), and \ref{fig:ss-temp}(c), respectively. In Figure \ref{fig:ss-temp}(a), we can see that the $\sigma_C$ tends to decrease linearly with increasing temperature, regardless of the nanotube chirality. $\varepsilon_C$ is almost temperature-independent for $(0,n)$-PHNTs, while decreases with increasing temperature for $(n,0)$-PHNTs, as shown in Figure \ref{fig:ss-temp}(b). The US values almost coincide with $\sigma_C$, i.e., they tend to decrease linearly with increasing temperatures, regardless of the nanotube chirality. At high-temperature regimes, US values tend to be similar for both tube chiralities.  

\begin{figure}[htb!]
    \centering
    \includegraphics[scale=0.5]{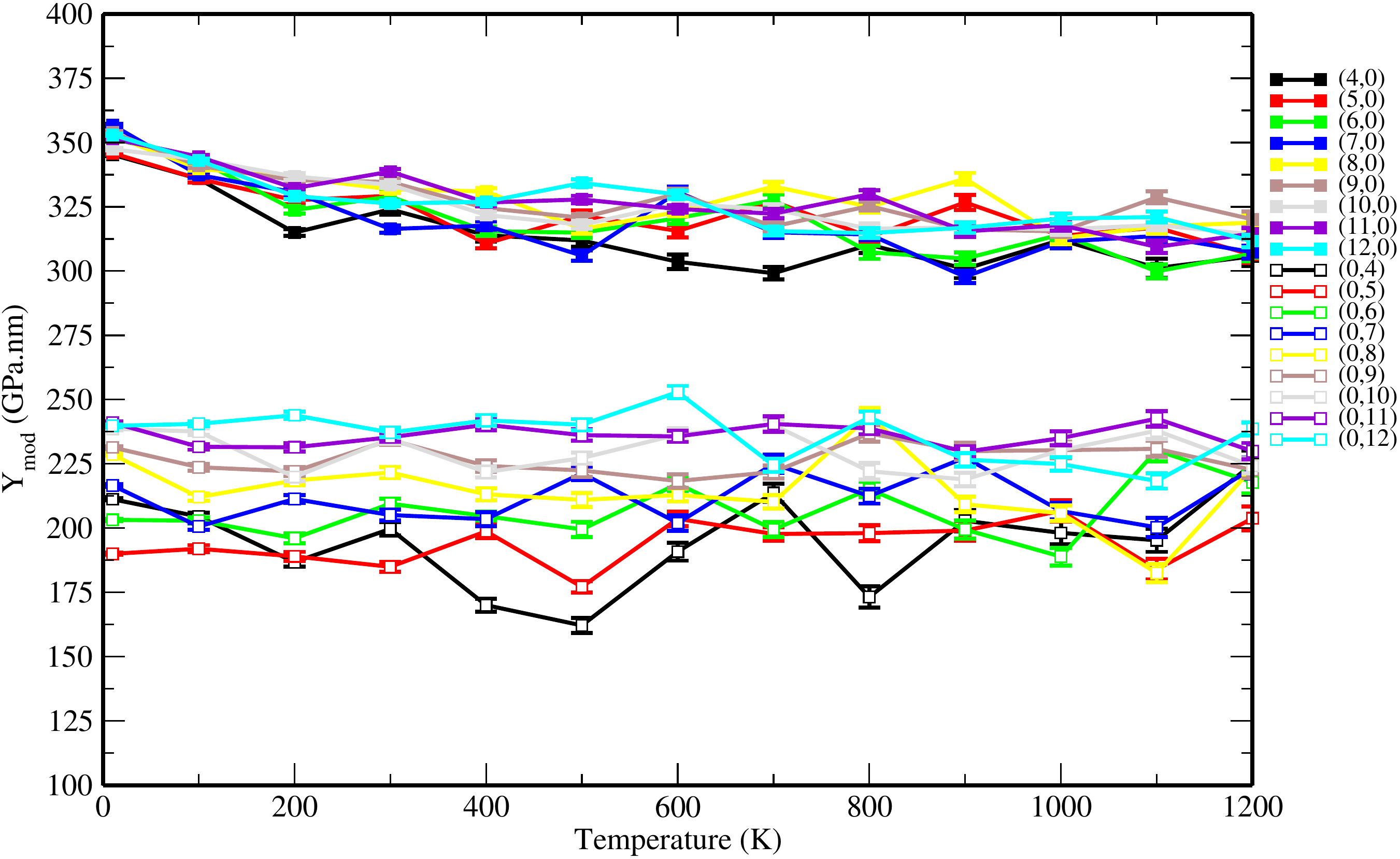}
    \caption{Young's modulus as function of temperature for all PHNTs investigated here.}
    \label{fig:ym-phnts}
\end{figure}

Young's modulus and Poisson's coefficients of the PHNTs were also calculated. Young's moduli were calculated from the slope of the stress-strain curves for the linear regime (up to 5 \% of strain). Their values do not change substantially for a particular PHNT (see Figure \ref{fig:ym-phnts}). Figure \ref{fig:ym-phnts} also shows that Young's modulus of $(n,0)$-PHNTs are about 150 GPa.nm, larger than that of $(0,n)$-PHNTs independent of the temperature. Although we have mentioned that $(n,0)$-PHNTs have more aligned C$-$C bonds and thus, they are able to accumulate larger strain than $(0,n)$-PHNTs, because their diameter are also larger than that of the $(0,n)$-PHNTs, the whole PHNT can be considered as modeled by a parallel association of a series of springs. The larger the diameter, the larger the number of parallel branches. As it is well known that a parallel association of springs is stiffer than a serial one, this might explain why the Young moduli of $(n,0)$-PHNTs are always larger than that of $(0,n)$-PHNTs. In fact, mechanical models based on the association of springs have been successfully used to describe the elastic properties of graphynes~\cite{cranford2012,kanegae2022}.

\begin{figure}[htb!]
    \centering
    \includegraphics[scale=0.5]{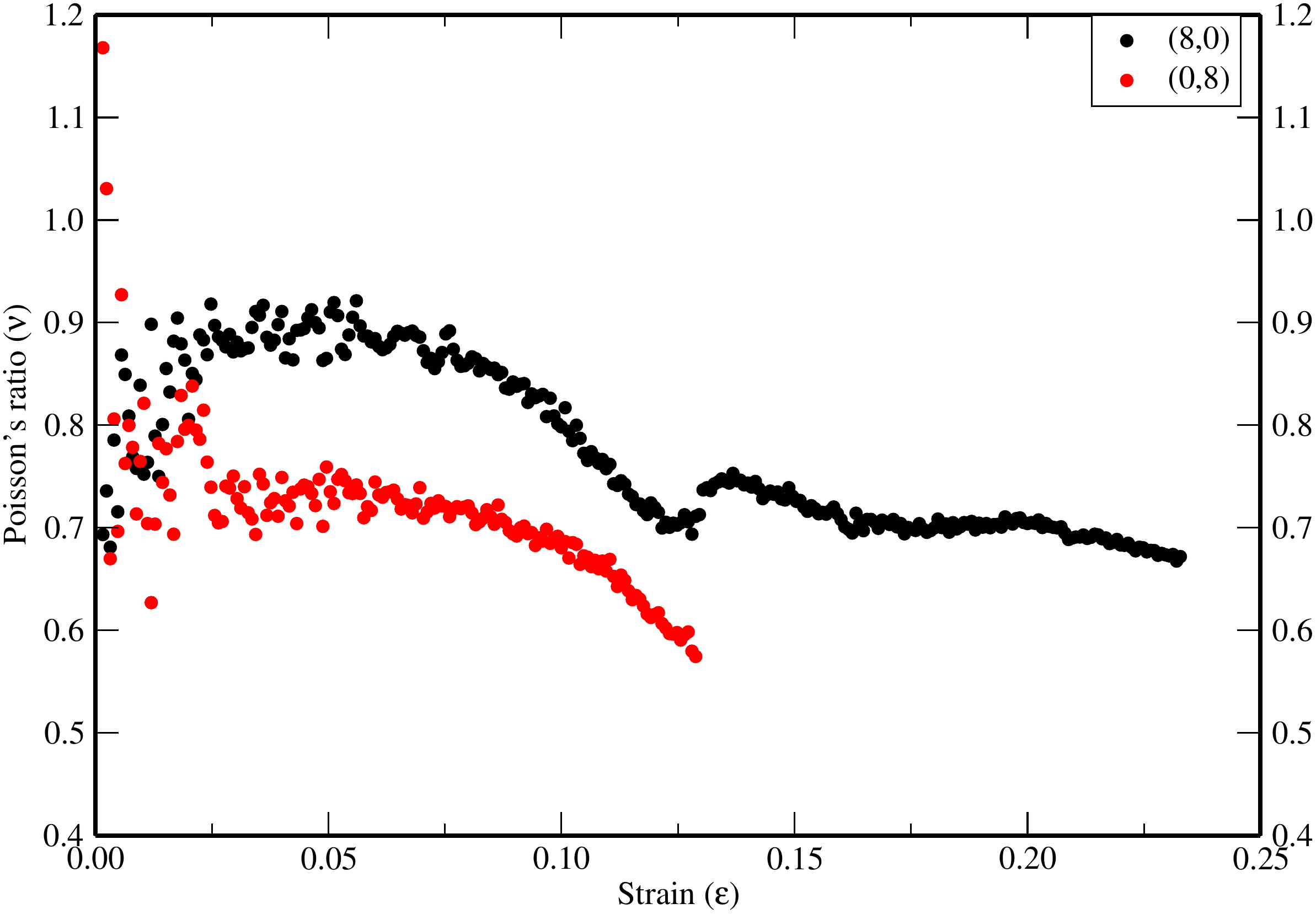}
    \caption{Poisson's ratio \textit{versus} strain for PHNTs (8,0) and (0,8) at room temperature.}
    \label{fig:poisson-coef}
\end{figure}

Contrary to previous predictions~\cite{sharma2014pentahexoctite}, our results for the Poisson's coefficient of $(n,0)$ and $(0,n)$-PHNTs are larger than 0.5. This indicates that PHNTs are stretch-densified materials, i.e., materials for which their volume decrease under tensile strain~\cite{rayfonseca2016}. As a consequence, they have negative linear compressibility along their axis direction~\cite{raysocrates1998,evans2015}. Figure \ref{fig:poisson-coef} shows the Poisson's coefficients of $(8,0)$-PHNT and $(0,8)$-PHNT as a function of the applied strain. It shows that the values of the  $(8,0)$-PHNT are always larger than that of the $(0,8)$-PHNT. This result is consistent with the fact that the deformation of the 8-membered rings during the tensile strain of the $(n,0)$-PHNTs is larger than that of the 6-membered rings of the $(0,n)$-PHNTs.

\section{Conclusions}

In this work, the mechanical properties (elastic and fracture patterns) of pentahexoctite nanotubes (PHNTs) were calculated for different values of temperature. Both reactive (ReaxFF) MD simulations and DFT calculations were pegrformed. The obtained values of Young's modulus,  ranging between 190-354 GPa.nm, agree very well with previously predicted values~\cite{sharma2014pentahexoctite}. The topology-dependent mechanical properties of PHNTs were also confirmed. When subjected to a critical strain, the PHNTs undergo an abrupt structural transiton to completely fractured states. In other words, there is no plastic regime when PHNTs are tensile strained. This might be a useful property towards the development of sensors or mechanical energy storage nanodevices. 

The mechanical properties of PHNTs were also studied as a function of temperature. We showed that the critical stress of PHNTs tends to decrease linearly with increasing temperature, regardless of the nanotube chirality. The critical strain values are almost temperature-independent for (0,$n$)-PHNTs and decrease with increasing temperature for ($n$,0)-PHNTs. Values of Young's modulus do not change substantially with $n$ for the ($n$,0)-PHNTs, but are always larger than that of (0,$n$)-PHNTs. ($n$,0)-PHNTs presented higher mechanical resilience than (0,$n$)-PHNTs due to the large number of C-C bonds along the tensile direction of the first. Also, the octagon ring can accumulate more strain than the hexagon rings in both structures when tensile strained. US tends to decrease linearly with increasing temperature, regardless of the nanotube chirality. At high-temperature regimes, US values tend to be similar for both tube chiralities.  

\section{Acknowledgements}

Authors acknowledge funding support from the following Brazilian Agencies: CAPES, CNPq, FAPDF, FAPESP and FAPEPI. W.H.S.B., A.L.A. and J.M.S thank the Laboratório de Simulação Computacional Cajuína (LSCC) at Universidade Federal do Piau\'i for computational support. J.M.S and D.S.G thank the Center for Computational Engineering and Sciences at Unicamp for financial support through the FAPESP/CEPID Grant \#2013/08293-7. J.M.S. and L.A.R.J acknowledge CENAPAD-SP (Centro Nacional de Alto Desenpenho em São Paulo - Universidade Estadual de Campinas - UNICAMP) for computational support process (proj634 and proj842). A.L.A. acknowledges CNPq (Process No. $427175/20160$) for financial support.  L.A.R.J acknowledges the financial support from a Brazilian Research Council FAP-DF grants $00193-00000857/2021-14$, $00193-00000853/2021-28$, and $00193-00000811/2021-97$ and CNPq grant $302236/2018-0$ and $350176/2022-1$, respectively. L.A.R.J. also gratefully acknowledges the support from ABIN grant 08/2019 and N\'ucleo de Computaç\~ao de Alto Desempenho (NACAD) for providing the computational facilities through the Lobo Carneiro supercomputer. L.A.R.J. thanks Fundaç\~ao de Apoio \`a Pesquisa (FUNAPE), Edital 02/2022 - Formul\'ario de Inscriç\~ao N.4, for the financial support. A.F.F thanks the Brazilian Agency CNPq for Grant No. 303284/2021-8 and São Paulo Research Foundation (FAPESP) for Grant No. \#2020/02044-9. This research also used the computing resources and assistance of the John David Rogers Computing Center (CCJDR) in the Institute of Physics Gleb Wataghin, University of Campinas.

\newpage

 \bibliographystyle{elsarticle-num} 
 \bibliography{PHXNT.bib}
\end{document}